\documentclass{iopart}
\usepackage{graphicx}% Include figure files
\usepackage{bbold}% Bold math
\usepackage{color}% Color text
\usepackage[english]{babel}
\usepackage{iopams} %Modified versions(?) of amsgen, amsfonts, amsbsy, amssymb
\usepackage{setstack} %11 macros from AMSmath
\newenvironment{align}{\begin{eqnarray}}{\end{eqnarray}}
\newenvironment{align*}{\begin{eqnarray*}}{\end{eqnarray*}}
\newcommand\eqref[1]{(\ref{#1})}
\newenvironment{bmatrix}{\left [\!\! \begin{array}{cc}}{\end{array}\!\! \right ]}
\newenvironment{b1matrix}{\left [\!\! \begin{array}{c}}{\end{array}\!\! \right ]}
\newenvironment{b3matrix}{\left [\!\! \begin{array}{ccc}}{\end{array}\!\! \right ]}
\newenvironment{qmatrix}{\left (\!\! \begin{array}{cc}}{\end{array}\!\! \right )}

\newenvironment{widetext}{}{}

\renewcommand{\phi}{\varphi}
\renewcommand{\epsilon}{\varepsilon}

\newcommand\op[1]{\hat{#1}} %Operator
\newcommand\wek[1]{\vec{#1}} %Makes it easy to change style. 3D vectors
\newcommand\vek[1]{\mathbf{#1}} %2D vectors are bold face.

\newcommand\abs[1]{\parallel #1 \parallel}

\newcommand\Ide{{\bf 1}} %Matrix

\newcommand\kv{\vek{k}}
\newcommand\kpv{\vek{k}'}
\newcommand\kppv{\vek{k}''}
\newcommand\rv{\vek{r}}
\newcommand\rpv{\vek{r}'}
\newcommand\qv{\vek{q}}

\newcommand\rw{\wek{r}}
\newcommand\pw{\wek{p}} %3D vectors are indicated by \wek

\newcommand\cf{\chi^f}

\newcommand\kfo{k_F^{0}}
\newcommand\ntd{n_{{\rm 2D}}}
\newcommand\Vex{V^{{\rm ext}}}

\newcommand\csch{{\rm csch}}
\newcommand\sech{{\rm sech}}
\newcommand\erfc{{\rm erfc}}

\newcommand\Hop{\op{H}}
\newcommand\rop[1]{\op{\rho}_{#1}}
\newcommand\co[1]{\op{c}_{#1}}
\newcommand\cd[1]{\op{c}^{\dagger}_{#1}}

\begin{document}

\title{Density wave instabilities of tilted fermionic dipoles in a multilayer geometry}

\author{J K Block, N T Zinner and G M Bruun}
\address{Department of Physics and Astronomy, University of Aarhus, DK-8000 Aarhus C, Denmark}
\ead{jkblock@phys.au.dk}

\date{\today}

\begin{abstract}
We consider the density wave instability of fermionic dipoles aligned by an external field, and moving in equidistant layers at zero temperature. 
Using a conserving Hartree-Fock approximation, we show that correlations between dipoles in different layers significantly decrease the critical coupling strength
for the formation of density waves when the distance between the layers is comparable to the inter-particle distance within each layer. 
This effect, which is strongest when the dipoles 
are oriented perpendicular to the planes, causes the density waves in neighboring layers to be in-phase for all orientations of the
dipoles. We furthermore demonstrate that the effects of the interlayer interaction can be understood from a classical model. Finally, we 
show that the interlayer correlations are important for experimentally relevant dipolar molecules, including the chemically stable $^{23}$Na$^{40}$K and $^{40}$K$^{133}$Cs, where the density wave regime is within experimental reach.

\end{abstract}

\pacs{03.75.Ss,67.85.−d,68.65.Ac,73.20.Mf}% PACS, the Physics and Astronomy
                             % Classification Scheme.

\maketitle

\section{Introduction}
The trapping and cooling of molecules in their rotational and vibrational ground state is a new research
direction within the field of ultracold atomic and molecular 
physics \cite{ospelkaus2008,ni2008,deiglmayr2008,lang2008,ni2010,ospelkaus2010,aikawa2010,danzl2010,miranda2011,chotia2011}. 
In contrast to the short-range isotropic interactions in typical cold atomic gases, molecules provide 
anisotropic potentials that typically have a long range dipolar part. 
This opens up a host of possibilities for exploring interesting physics \cite{carr2009,baranov2008,lahaye2009}
and also chemical reaction dynamics at low temperatures \cite{krems2008,hutson2010}.

Chemical reaction losses can be large in three-dimensional (3D) samples \cite{ospelkaus2010} due to the attractive 
head-to-tail interaction of dipolar molecules. However, recent experiments using optical lattices \cite{miranda2011,chotia2011}
have shown that a low-dimensional confinement of the system can effectively suppress the loss. Interesting 
many-body phases have been proposed in such settings, including $s$- and $p$-wave superfluid states in single- and
bilayer setups \cite{Bruun,cooper2009,sarma2009,pikovski2010,zinner2010,baranov2011,levinsen2011,ticknor2011}, 
density-waves in homogeneous \cite{SunPhysRevB.82.075105,Yamaguchi,Zinner,BabadiPRB.84.235124,Sieberer,parish2011}
and lattice systems \cite{Mikelsons2011,Bhongale2011,Gadsbolle2012},
and non-trivial Fermi liquid behavior \cite{chan2010,kestner2010,lu2011}. The long-range dipolar forces 
also opens up for a very rich spectrum of few-body bound state physics in 1D \cite{deu2010,santos2010,wunsch2011,zollner2011} and 
2D setups \cite{klawunn2010a,cremon2010,artem2011b,armstrong2011a,artem2011c}.
\begin{figure}[htb]
\begin{center}
 \includegraphics[width=0.49\textwidth,keepaspectratio=1]{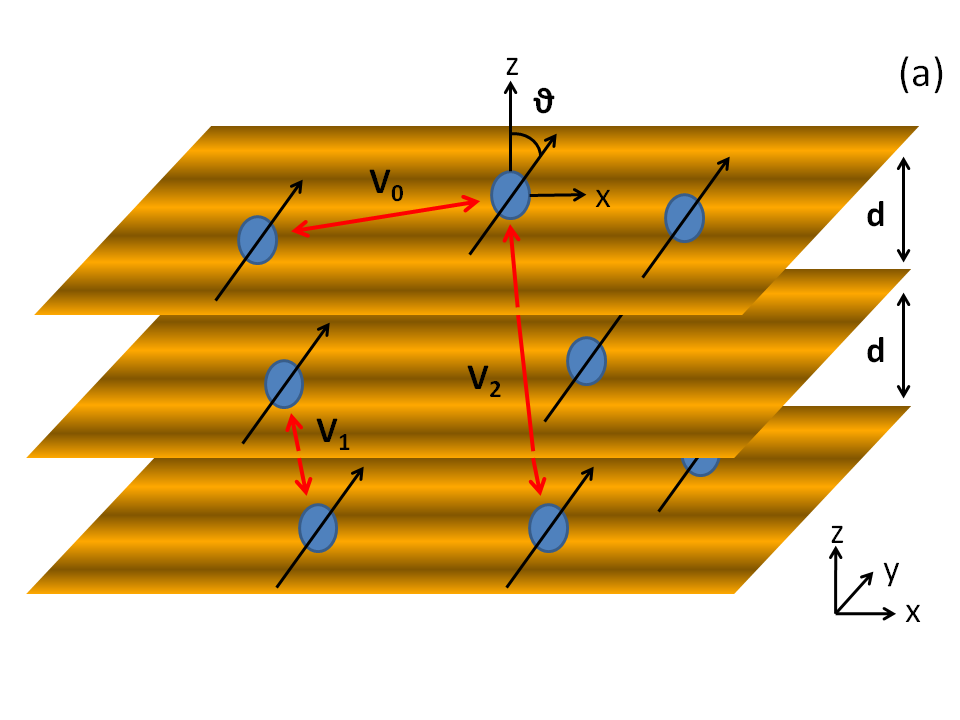}
\includegraphics[width=0.49\textwidth,keepaspectratio=1]{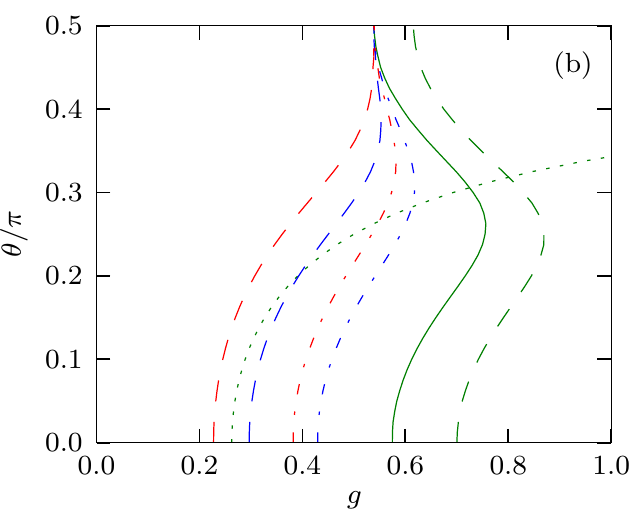}
\caption{(a) Illustration of the setup considered. The dipoles reside in layers in the $xy$-plane separated by the distance $d$. The angle $\theta$, between the dipoles and the normal to the layers is in the $xz$-plane. The interaction between dipoles in the same layer, in adjacent layers, and in layers separated by a
distance $2d$ is indicated by $V_0$, $V_1$, and $V_2$ respectively. 
(b) The phase boundary between the normal (left) and the striped phase (right) in the $(g,\theta)$ plane for strictly 2D layers with $w=0$: 
 a single layer (\textcolor{green}{$\full$}), two layers separated by $d\kfo=0.5$ (\textcolor{blue}{$\dashed$}) and $d\kfo=1$ (\textcolor{blue}{$\chain$}), and three layers separated by $d\kfo=0.5$ (\textcolor{red}{$\dashed$}) and $d\kfo=1$ (\textcolor{red}{$\chain$}). The \textcolor{green}{$\dotted$} line gives the RPA (Random Phase Approximation) result for a single layer for comparison. The case of a single quasi-2D layer with $w \kfo=0.1$
 is plotted by \textcolor{green}{$\dashed$}.
} \label{fig:setup} 
% \label{fig:birelsingle}
\end{center}
\end{figure}

We consider the density wave (stripe) instabilities of fermionic dipoles at zero temperature in 2D layers. The dipole moments
are aligned by an external field, and they are moving in
equidistant layers as illustrated in figure~\ref{fig:setup}(a). Calculating the density-density response function within a conserving 
Hartree-Fock approximation, we find the following. %(I) Exchange correlations increase the critical value of the coupling strength for the formation of stripes significantly as compared to predictions based on the random-phase-approximation (RPA). They also affect the shape of the phase boundary between the normal and the striped phase as a function of the angle $\theta$ in a qualitative way. 
(I) The presence of several layers can decrease the critical coupling strength for stripe formation significantly. The effect is strongest for the dipoles oriented 
perpendicular to the planes whereas it vanishes when the dipoles are aligned in the planes, see figure~\ref{fig:setup}(b).
(II) The mechanism for this decrease is the formation of stripes on top of each other in adjacent layers. This in-phase stripe pattern is always energetically favorable, 
independent of the orientation of the dipole moment which is somewhat surprising. It can be understood from a purely classical calculation which also explains the 
angle dependence of the effect. 
\begin{figure}[ht]
\begin{center}
\includegraphics[width=0.9\textwidth,keepaspectratio=1]{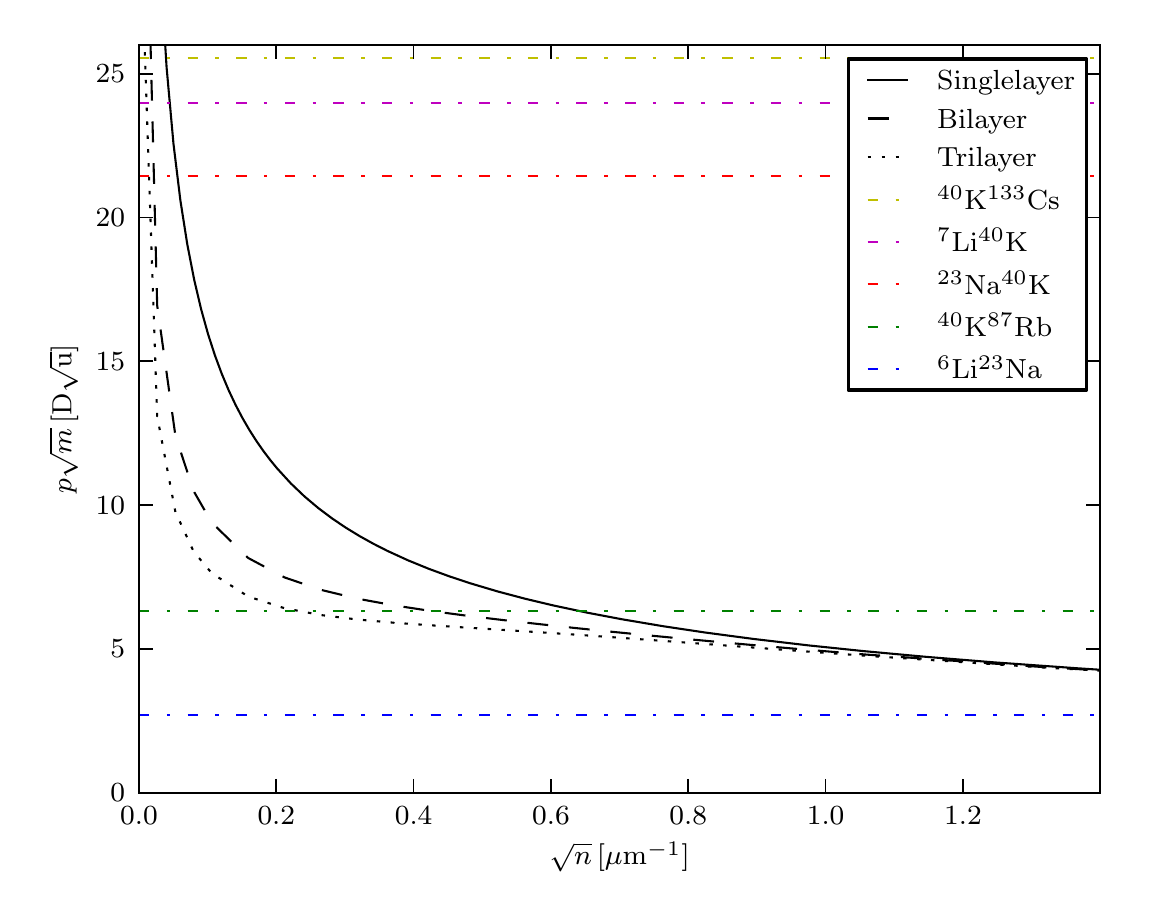}%
\caption{For a typical layer separation $d=1064\,\mathrm{nm}/2$ and perpendicular polarization $\theta=0$, the critical value of dipole moment times the square root of 
the mass, $p\sqrt{m}$, as a function of the square root of the two dimensional density for the single-, bi- and trilayer geometries. The horizontal lines indicate the 
permanent electrical dipole moment of the ground state (values from \cite{igel-mann1986}) times the square root of the mass for the respective molecules. Outside the 
range of the axis lie $^{6}$Li$^{87}$Rb and $^{6}$Li$^{133}$Cs with $p\sqrt{m}$ of 40.0 $\mathrm{D}\sqrt{\mathrm{u}}$ and 64.6 $\mathrm{D}\sqrt{\mathrm{u}}$ 
respectively.}
 \label{fig:ExpFig}
 \end{center}
\end{figure}
(III) The decrease in the critical coupling strength for stripe formation due to interlayer correlations is significant for experimentally relevant systems consisting of 
$^{7}$Li$^{40}$K, $^{23}$Na$^{40}$K, $^{40}$K$^{133}$Cs, $^{6}$Li$^{87}$Rb and $^{6}$Li$^{133}$Cs molecules. It is less important 
for $^{40}$K$^{87}$Rb and $^6$Li$^{23}$Na molecules which have smaller dipole moments.

\section{System}
The system consists of identical fermionic dipoles of mass $m$ moving in equidistant layers separated by a distance $d$. Their dipole moment $\pw$ is aligned 
forming the angle $\theta$ with respect to the normal of the layers ($z$-axis) with their projection onto the planes defining the $x$-axis, see figure~\ref{fig:setup}. The layers are formed by a deep 1D optical lattice so that the dipoles in layer $l$ ($l$ an integer) reside in the lowest Wannier state in the z-direction. This is to a good approximation a Gaussian 
$\phi_l(z)=\exp[-(z-ld)^2/2w^2]\pi^{-1/4}w^{-1/2}$, where $w$ is the width of the layer which is centered at $ld$. We neglect any trapping potential in the $xy$-plane so that the transverse states are labelled by the momentum $\kv=(k_x,k_y)$ (we use units where $\hbar=k_B=1$). 

In this basis the grand canonical Hamiltonian reads
\begin{equation}
\fl\Hop=\sum_{\kv,l}\bigg(\frac{k^2}{2m}-\mu\bigg)\cd{\kv,l}\co{\kv,l}+\frac{1}{2}\sum_{l,l'}\sum_{\kv,\kpv,\qv}V_{l,l'}(\qv)\cd{\kv+\qv,l}\cd{\kpv-\qv,l'}\co{\kpv,l'}\co{\kv,l},
 \label{eq:Hamiltonian}
\end{equation}
where we have absorbed the harmonic oscillator energy of the $z$-direction 
 in the chemical potential $\mu$. Here, $\hat c_{\kv,l}$ removes a dipole in layer $l$ with momentum $\kv$.
The interaction between two dipoles separated by $\rw$ is 
\begin{equation}
V_{\rm 3D}(\rw)=D^2\frac{1-3\cos^2(\theta_r)}{r^3}
\label{eq:Vrealspace}
\end{equation}
where $\theta_r$ is the angle between $\rw$ and the dipole moment $\pw$, and $D^2=p^2/4\pi\epsilon_0$
for electric dipoles. The effective interaction $V_{l,l'}(\qv)$ 
between dipoles in layer $l$ and $l'$ is obtained by integrating the interaction $V_{\rm 3D}(\rw)$ over the 
 Gaussians $\phi_l(z)\phi_{l'}(z)$ 
combined with a 2D Fourier transform. This yields~\cite{fischer2006} 
\begin{equation}
V_0(\qv)=\pi D^2\bigg [\frac{8}{3w\sqrt{2\pi}} P_2(\cos\theta)-2\xi(\theta,\phi)F(q)\bigg ].
\label{eq:V2Dq}
\end{equation}
for the intralayer interaction where $P_2(x)=(3x^2-1)/2$ is the second Legendre polynomial, while 
\begin{equation}
\fl F(q)= q\exp[(qw)^2/2]\erfc[qw/\sqrt{2}] \quad \mathrm{and} \quad \xi(\theta,\phi)=\cos(\theta)^2-\sin(\theta)^2\cos(\phi)^2 \label{eq:xi}.
\end{equation}
The constant term in equation (\ref{eq:V2Dq}) corresponds to a $\delta(\rv)$ interaction which plays no role since we consider identical fermions. 
The real part of the interlayer interaction is only dependent on the difference $l$ in layer numbers and is given by~\cite{Wang,Qiuzi}
\begin{equation}
 \label{eq:Vinter}
V_l(\qv)=-2\pi D^2\xi(\theta,\phi)q e^{-dl q}
\end{equation}
for $w\ll d$. This approximation deviates less than $10\%$ from the exact expression for $w\le d/5$. In equation (\ref{eq:Vinter}), we have only given the real 
part of the potential since the imaginary part is zero for momenta along the $y$-direction which are the ones relevant for stripe formation, as discussed in section \ref{Sec:LinearResponse}.

The strength of the interaction is parametrized by the dimensionless number 
\begin{equation}
 \label{eq:gdef}
 g=\frac{4mD^2\kfo}{3\pi\hbar^2},
\end{equation}
 where $\kfo=\sqrt{4\pi n_{2D}}$ is defined from the 2D density. Likewise, a dimensionless measure for the layer separation is given by $d\kfo$.
 The ratio of the layer distance to the typical inter-particle distance in a layer is $d\kfo/\sqrt{4\pi}$.

\section{Linear response and the Hartree-Fock approximation}\label{Sec:LinearResponse}
To analyze the instabilities of the homogeneous phase towards the formation of stripes, we consider the retarded
density-density response function 
$\chi_{ij}^R(\rv-\rpv,t-t')=-i\theta(t-t')\langle \big [\rop{i}(\rv,t),\rop{j}(\rpv,t')\big]\rangle$
where $(i,j)$ denotes the layers and $\rv=(x,y)$. The density operator for layer
 $i$ is $\rop{i}(\rv)=\hat\psi_i^\dagger(\rv)\hat\psi_i(\rv)$ with $\hat\psi_i(\rv)$ 
the field operator for the dipoles in layer $i$. 
An instability toward the formation of a density wave with wave number ${\qv}$ shows up as a zero frequency ($\omega=0$) pole 
of the Fourier transformed response function $\chi_{ij}^R(\qv,\omega)$.

\begin{figure}[h]
 \centering
 \includegraphics[scale=0.5]{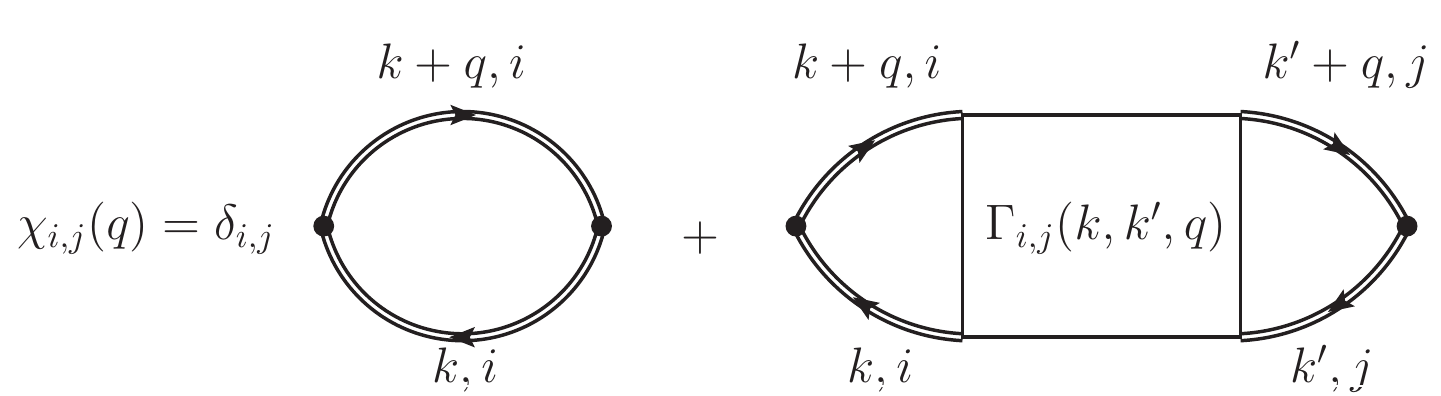}
\includegraphics[scale=0.5]{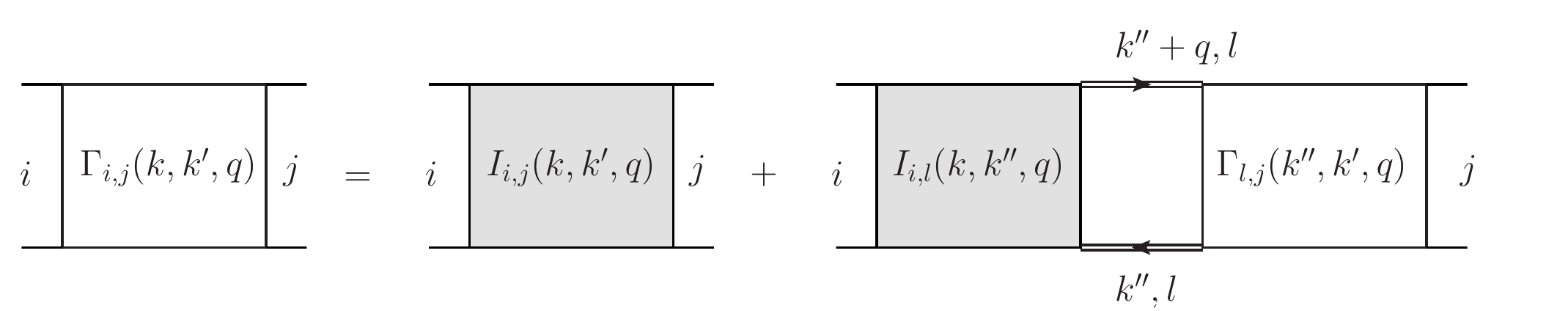}
  \caption{Top: Density-density response function. The thick lines indicate interacting Green's functions. Bottom: The Bethe-Salpeter equation for the particle-hole scattering matrix $\Gamma$. }
 \label{fig:ChiDiagram}
\end{figure}

We calculate the retarded response function using diagrammatic perturbation theory in Matsubara space $\chi_{ij}(\qv_c,i\omega_n)$
with $\omega_n=2n\pi T$ ($n$ integer) a bosonic Matsubara frequency. The retarded function $\chi_{ij}^R(\qv_c,\omega)$ is then 
obtained by analytical continuation $i\omega_n\rightarrow \omega+i0^{+}$ in the usual way~\cite{BruusFlensberg}.
 As illustrated in figure \ref{fig:ChiDiagram}, $\chi_{ij}(q)$ with $q=({\qv},i\omega_n)$ can be written as 
\begin{equation}
\label{eq:densityresponse}
 \chi_{ij}(q)=\sum_{k,k'}\big[\delta_{i,j}\delta_{k,k'}\Pi_i(k,q)+\Pi_i(k,q)\Gamma_{ij}(k,k',q)\Pi_j(k',q)\big]
\end{equation}
where $\Pi_i(k,q)=G_i(k+q)G_i(k)$ is the particle-hole propagator with $G_i(k)$ the single particle Green's function for the dipoles. 
The scattering matrix $\Gamma_{ij}(k,k',q)$, which describes a particle with momentum $k+q$ scattering on a hole with momentum $k$ 
to produce a particle with momentum $k'+q$ and a hole with momentum $k'$, obeys the Bethe-Salpeter equation 
\begin{equation}
 \label{eq:BS}
 \Gamma_{ij}(k,k',q)=I_{ij}(k,k',q)+\sum_{l,k''}I_{il}(k,k'',q)\Pi_{l}(k'',q) \Gamma_{lj}(k'',k',q)
\end{equation}
as shown in figure~\ref{fig:ChiDiagram}. Here $I_{il}(k,k'',q)$ includes all scattering processes which are 
irreducible with respect to the particle-hole propagator $\Pi_{l}(k,q)$. 

\begin{figure}[h]
 \centering
 \includegraphics[scale=0.5]{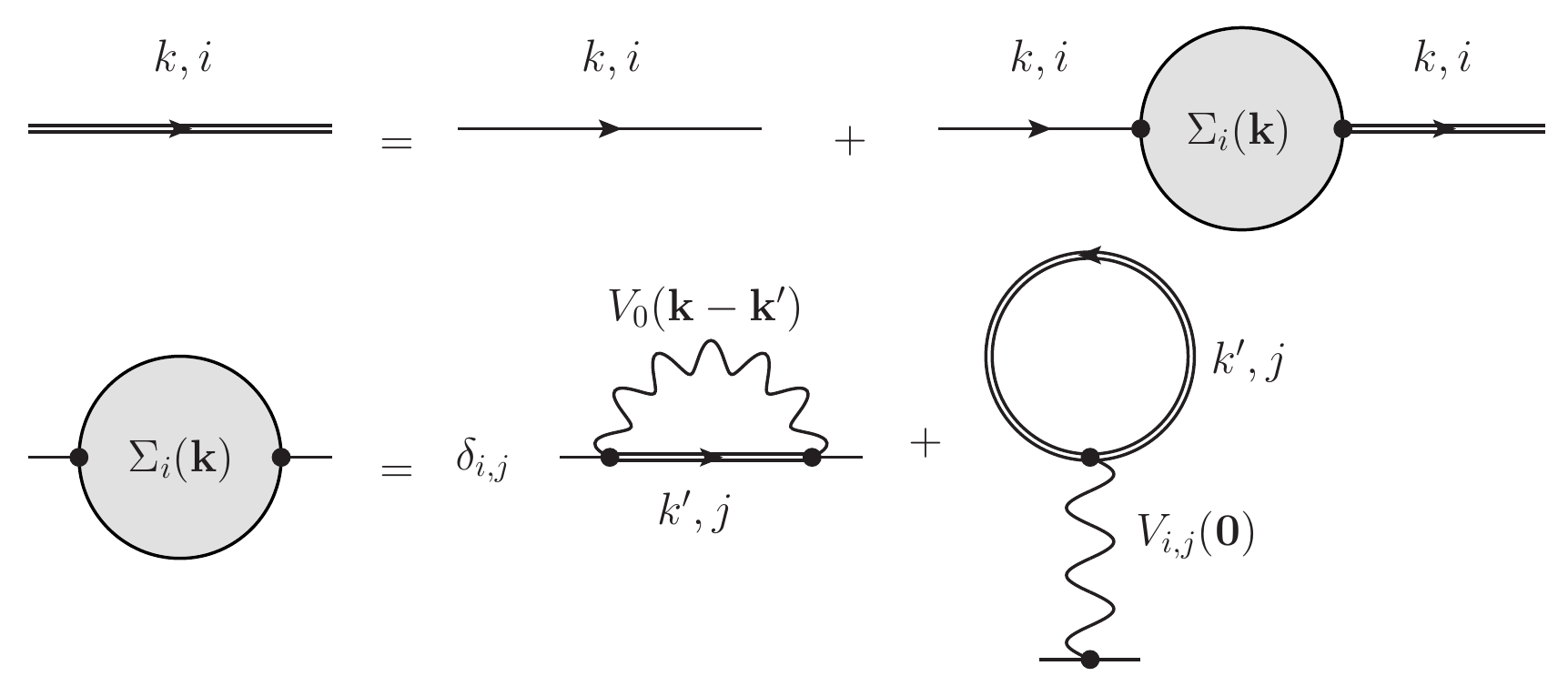}
\includegraphics[scale=0.5]{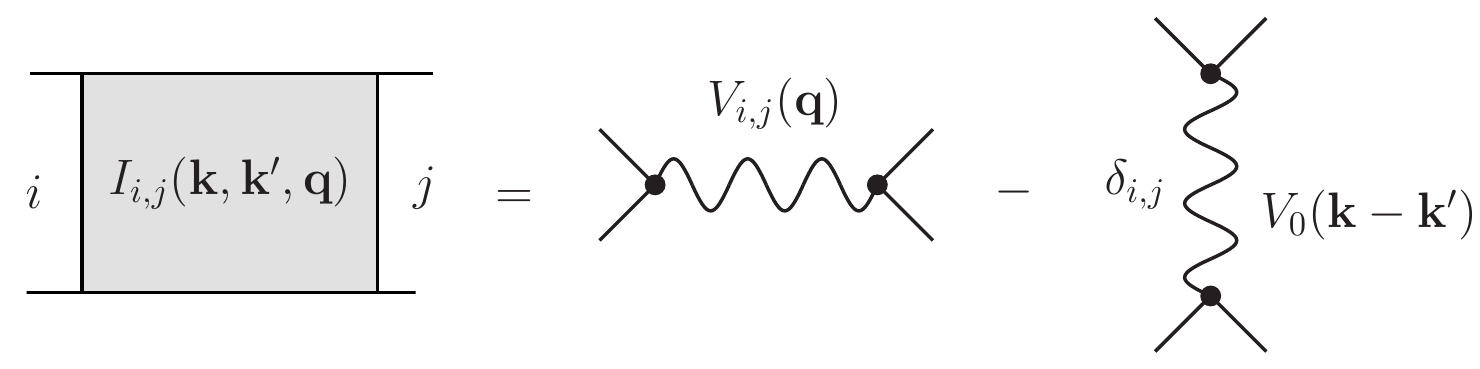}
 \caption{Hartree-Fock approximation for the Green's function and the irreducible particle-hole interaction. 
The thin lines indicate the non-interacting Green's functions and the wavy lines the interaction $V_{ij}$. }
 \label{fig:HFDiagrams}
\end{figure}

To proceed, we apply the self-consistent Hartree-Fock approximation illustrated in figure~\ref{fig:HFDiagrams}.
All Green's functions are interacting in this approximation and the vertex $I_{il}(k,k'',q)$ is given by the lowest order 
 direct and exchange interactions. 
Writing the Green's function as $G_l^{-1}(k)=ik_n-\epsilon_{l\kv}$ with $ik_n=(2n+1)\pi T$ a fermionic Matsubara frequency
and $\epsilon_{l\kv}=\kv^2/2m-\mu-\Sigma_l(\kv)$, we have 
the usual Hartree-Fock expression for the self-energy 
\begin{align}
\Sigma_l(\kv)=&\sum_{\kpv,j}[V_{l,j}(0)-\delta_{l,j}V_0(\kv-\kpv)]f_{l\kpv}. \label{eq:SE}
\end{align}
Here $f_{l\kv}=[\exp (\beta\epsilon_{l\kv})+1]^{-1}$ is the occupation of the $\kv$ momentum state in layer $l$.
In the calculations, we take $\beta=1/k_BT\rightarrow\infty$ appropriate
for the zero temperature case. It follows from (\ref{eq:Vinter}) that the 
 Hartree shift $V_{l,j}(0)$ for the energy of a dipole in layer $l$ due to the interaction with 
the dipoles in layer $j\neq l$ vanishes for thin layers, i.e.\ for $w\ll d$. 
 This is a consequence of the fact that the dipole-dipole interaction integrates to zero over a plane as shown in \ref{app:Classical}.
 In this paper, we keep $w\ll d$ and since 
 the density is the same in all layers, the self-energy and the chemical potential are both layer independent.
We calculate the single particle Green's functions numerically, obtaining self-consistency through an iterative 
procedure. The scattering matrix $\Gamma_{ij}(k,k',q)$ is then determined from equation (\ref{eq:BS}) using 
 \begin{equation}
 \label{eq:irredint}
I_{i,j}(\kv,\kpv,\qv) =V_{ij}(\qv)-\delta_{i,j}V_{ii}(\kv-\kpv).
\end{equation}
Note that exchange is only included for interactions within the same layer as dipoles in different layers are distinguishable. 
Diagrammatically, this approximation corresponds to the summation of bubbles containing 
 intralayer ladder interactions, which are connected to each other by inter- and intralayer interactions.

Since the irreducible interaction vertex $I_{i,j}$ is independent of frequency in this approximation, the particle-hole scattering matrix is independent of the internal frequencies $ik_n',ik_n''$ and we can perform 
these frequency summations in equation (\ref{eq:BS}), which then simplifies to 
\begin{equation}
 \label{eq:scat}
\fl \Gamma_{ij}(\kv,\kpv,q)=I_{ij}(\kv,\kpv,\qv)+\sum_{\kppv l} I_{il}(\kv,\kppv,\qv)
\frac{f_{l\kv''}-f_{l\kv''+\qv}}{iq_n+\epsilon_{l\kv''}-\epsilon_{l\kv''+\qv}}
 \Gamma_{lj}(\kppv,\kpv,q).
\end{equation}
Equation \eqref{eq:scat} corresponds to a series of inhomogeneous Fredholm equations of the second kind in the first variable.
From equation (\ref{eq:densityresponse}) it follows that apart from the poles of $\Pi$ describing the particle-hole continuum, the density-density response function $\chi$ and the particle-hole scattering matrix $\Gamma$ share the same poles describing collective modes. Thus, we determine the critical coupling strength by searching for a zero frequency pole at a non-zero momentum $\qv$ of the matrix $\Gamma_{ij}(\kv,\kpv,q)$, which is analytically continued to real frequency by $iq_n\to \omega + i0^{+}$.

When $\theta>0$, the anisotropic interaction favors dipoles that are aligned along the $x$-axis. This corresponds to a density wave with a wave vector pointing in the perpendicular direction $\phi_c=\pi/2$. Since the density wave is formed by particle-hole excitations, we expect the length of the wave vector to be $2k_F(\phi_c)$ to minimize the kinetic energy cost. This indeed follows directly from RPA calculations of the density-density response function \cite{SunPhysRevB.82.075105,Yamaguchi}, and we expect it to hold even 
 when exchange correlations are included. 
 
The self-consistent Hartree-Fock calculation of the response function is demanding numerically, and we describe in \ref{app:Numerics} how it is implemented numerically. The payoff is that the approximation is conserving in the sense of Kadanoff-Baym \cite{baym1961}. We shall furthermore see that the exchange correlations have large effects on the critical coupling strength for the formation of stripes.

\section{Numerical results}
We now present numerical results for the critical coupling strength $g_c$ as a function of dipole angle $\theta$, layer separation $d\kfo$, and layer thickness $w\kfo$ for fixed momentum in the direction $\phi_c=\pi/2$ with magnitude $q=2k_F(\phi_c)$. We shall for concreteness consider the cases of one, two, and three layers.

\subsection{Single layer}
We first focus on the case of a single layer with vanishing thickness, i.e.\ $w=0$. In figure~\ref{fig:setup}(b), we plot as a solid line the phase boundary between the normal phase (left) and the striped phase (right)
for a single layer. The boundary has an intriguing non-monotonous behavior with a maximum critical coupling strength $g_c$
for $\theta\simeq \pi/4$. For comparison we also plot
as a dotted line the result of a RPA calculation using the interacting Green's functions~\cite{SunPhysRevB.82.075105,Yamaguchi}.

For small dipole angles $\theta$, the RPA result underestimates the critical coupling strength significantly as compared 
to the conserving HF approximation. This demonstrates that 
exchange correlations suppress the formation of stripes. 
 For larger angles, the shape of the phase boundary differs qualitatively from that obtained from the RPA calculation which predicts 
 $g_s\propto \cos(\theta)^2$ (neglecting the effects of Fermi surface deformation). 

The dependency on $\theta$ for the RPA result (green, dotted line of figure~\ref{fig:setup}(b)) can be understood purely from the fact that the repulsive part 
of the interaction decreases as $\cos^2\theta$ as the dipoles are tilted towards the layer. For small $\theta$, the exchange correlations suppress stripe formation. As $\theta$ crosses the ``magic angle'' $\cos^{-1}(1/\sqrt{3})$, the spatial 
(or momentum) average of the interaction goes from being repulsive to being attractive. 
Since exchange correlations enter through an average over the momentum transfer in the term $V(\qv)-V(\kv-\kpv)$, see \eqref{eq:irredint} and the $k'$-sum in \eqref{eq:BS}, this means that the effects of the exchange term vanishes right at the magic angle as can be seen from figure~\ref{fig:setup}(b). For larger $\theta$, exchange correlations enhance the stripe instability and for $\theta\rightarrow \pi/2$ the instability is entirely driven by the exchange term 
 since the direct interaction vanishes. This is the qualitative origin of the maximum in the critical coupling strength for an angle close to $\cos^{-1}(1/\sqrt{3})$.

 Our result for the phase boundary agrees within $7\%$ with that of ref.~\cite{Sieberer}, and the critical coupling strength for $\theta=0$ agrees within $5\%$ of that reported in ref.~\cite{BabadiPRB.84.235124}.
 As explained in \ref{app:Numerics}, we have taken care to use many $k$-space points in the integration around the singular points in (\ref{eq:scat}),
 and we estimate our results to be numerically accurate within $1 \%$. On the other hand, our results differ substantially from those obtained 
 using a self-consistently determined local field factor~\cite{parish2011}.

 We also plot in figure~\ref{fig:setup}(b) the phase boundary for a quasi-$2D$ single layer system where the finite depth of the 1D optical lattice gives a width of $w\kfo=0.1$ for the Gaussian transverse wavefunctions. This softening of the $z$-direction reduces the repulsive part of the effective dipole-dipole interaction (\ref{eq:V2Dq}), and as a result the critical coupling strength for stripe formation is increased by up to $18\%$. For larger $w\kfo$, one has to take into account higher states in the $z$-direction as investigated for the case of $\theta=0$ in Ref.~\cite{BabadiPRB.84.235124}.

Note that we for simplicity have not included the region of $p$-wave superfluidity for $\theta\gtrsim \pi/4 $~\cite{Bruun} in the phase diagram, nor the 
region of collapse due to a negative compressibility for large angles and coupling strengths~\cite{Bruun,Yamaguchi,parish2011}. 

\subsection{Several layers}
\begin{figure}[htb]
\begin{center}
\includegraphics[width=0.49\textwidth,keepaspectratio=1]{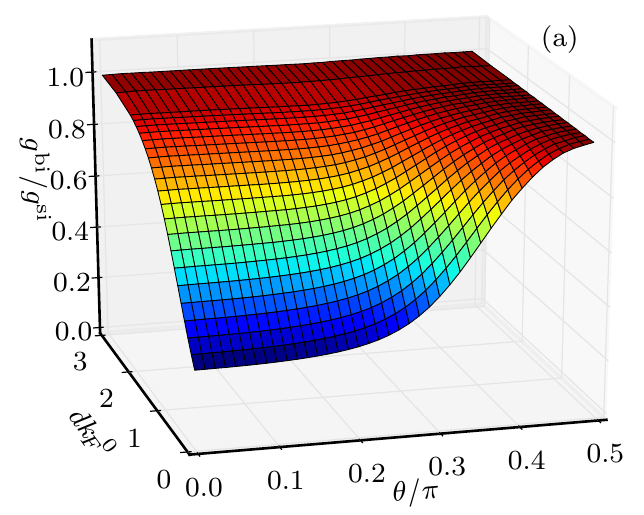}
\includegraphics[width=0.49\textwidth,keepaspectratio=1]{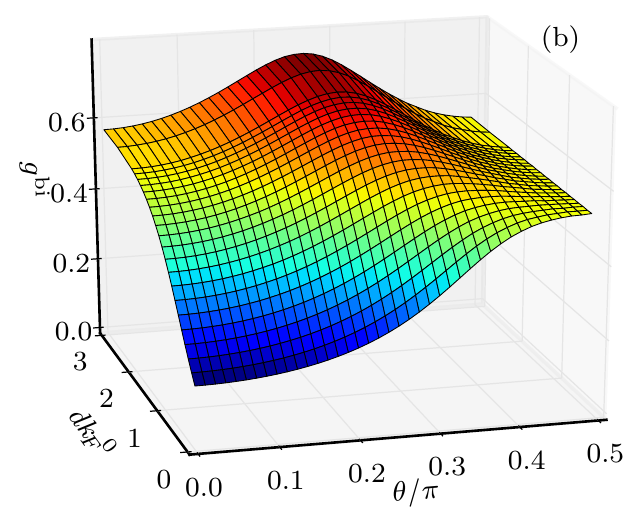}
\end{center}
 \caption{ The critical coupling strength $g^{\mathrm{bi}}$ for stripe formation for two layers as a function of dipole angle $\theta$ and 
 distance $d \kfo$ between the layers for $w=0$: (a) $g^{\rm bi}/g^{\rm si}$ and (b) $g^{\rm bi}$.} 
 \label{fig:bilayerdtheta}
\end{figure}

In figure \ref{fig:setup}(b), we also plot the phase boundary in the case of two and three layers separated by the distances 
$d\kfo=0.5$ and $d\kfo=1$. This illustrates a main result of this paper: the presence of neighbouring layers reduces 
the critical coupling strength for stripe formation significantly when the layer distance $d$ is comparable to the distance between particles within each layers. 
As expected, the effect decreases with increasing layer separation as follows directly from the exponential decay of the inter-layer interaction (\ref{eq:Vinter}). This is illustrated further in figure \ref{fig:bilayerdtheta}(a) where 
the critical coupling strength for the bi-layer case is compared to that of the single layer case as a function of layer separation and dipole angle. 
For small layer separation, the critical coupling strength can in fact be shown to scale as $1/N$ for $N\gg 1$ layers, since the exchange correlations within each 
layer can be neglected in this limit, so that the problem reduces to that of dipoles with $N$ internal degrees of freedom moving in a single 
layer~\cite{Zinner}.
 From figure \ref{fig:bilayerdtheta}(a) we furthermore conclude that the effects of neighboring layers is strongest for small dipole angles. 
 This effect has a simple classical interpretation as we shall demonstrate below. 

Figure \ref{fig:bilayerdtheta}(b) shows the critical coupling strength for the bi-layer case for $w=0$. It increases with increasing layer separation 
$d\kfo$ whereas it has an interesting angular dependence as a result of the interplay between the angular dependence of the 
interlayer and intralayer interaction. 
Note that for well-separated layers with $w\ll d$, a non-zero layer thickness only changes the intra-layer interaction,
and it therefore affects the critical coupling strength in a way similar to that of a single layer. 

It should be noted that $s$-wave interlayer superfluidity is likely to 
occur in the multilayer system. This has been discussed for perpendicular dipoles in for both bilayer \cite{pikovski2010,zinner2010,baranov2011}
 and multilayer cases (restricted to nearest-neighbour pairing) \cite{potter2010}.

For $\theta=0$, the interlayer superfluidity is expected for any value of $g$ when density-waves are ignored. At sufficiently large $g$, we will likely have a competition of the two phases and a model containing both is necessary to infer which phase dominates or whether there can be coexistence. 
The superfluid could have $p$-wave symmetry for larger angles $\theta$ which may be more favorable to coexistence. This will be explored in future studies.

\subsection{Correlations between stripes in different layers}
To examine further the effects of neighboring layers, 
we now analyze the $\omega=0$ collective mode and the associated density oscillations at the critical coupling strength $g_c$. We first 
focus on the case of two layers. The density oscillations induced by an external perturbation 
$\sum_l\int d^2r\Vex_l({\rv},t)\hat \rho_l({\rv})$ are within linear response given as 
\begin{align}
 \begin{bmatrix}
\delta \rho_1({\qv},\omega)\\
\delta \rho_2({\qv},\omega)\\
 \end{bmatrix}
=
\begin{bmatrix}
\chi_{1,1}({\qv},\omega) & \chi_{1,2}({\qv},\omega)\\
 \chi_{2,1}({\qv},\omega) & \chi_{2,2}({\qv},\omega)
 \end{bmatrix}
 \begin{bmatrix}
\Vex_1({\qv},\omega)\\
\Vex_2({\qv},\omega)
 \end{bmatrix}.
 \label{densityrespmatrix}
\end{align}
At the critical coupling strength $g_c$, one eigenvalue of the density response matrix in equation (\ref{densityrespmatrix}) diverges.
We find that the mode which first diverges always is symmetric in the layer index $l$, independent of the dipole angle, except 
 for $\theta=\pi/2$ where the modes are degenerate.
 Close to the critical coupling $1-g/g_c\ll 1$, the density-density response matrix has the pole structure 
 \begin{align}\label{eq:biresponsepole}
\begin{bmatrix}
\chi_{1,1} & \chi_{1,2}\\
 \chi_{2,1} & \chi_{2,2}
 \end{bmatrix}
=&
\begin{bmatrix}
\chi^0 & 0\\
0& \chi^0
\end{bmatrix}
+\frac{1}{1-g/g_c}
\begin{bmatrix}
\chi^c & \chi^c \\ 
\chi^c& \chi^c
 \end{bmatrix},
\end{align}
as we demonstrate in detail in \ref{app:nearcrit}. Here,
 $\chi^0$ is the Lindhard function including Hartree-Fock shifts of the single particle energies. 
Thus, the density instability in the two layers is in-phase and the stripes will be on top of each other as illustrated in 
figure (\ref{fig:setup}). The mode which is anti-symmetric in the layer index becomes unstable at a larger coupling strength.
This can be understood as a splitting of the eigen-mode for a single layer into a symmetric and anti-symmetric mode. 

\begin{figure}[htb]
\begin{center}
\includegraphics[width=0.49\columnwidth,keepaspectratio=1]{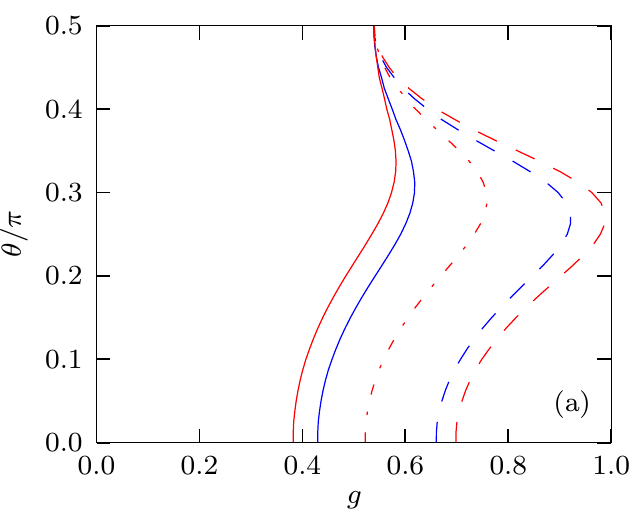}
\includegraphics[width=0.49\columnwidth]{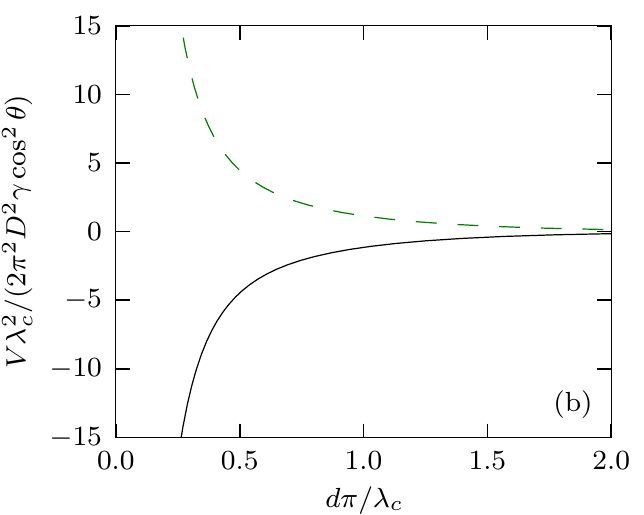}
\caption{(a) The lowest density waves modes for the layer separation $d\kfo=1$ and thickness $w=0$. Bilayer in-phase (\textcolor{blue}{$\full$}) and anti-phase (\textcolor{blue}{$\dashed$}), and trilayer all layers in-phase (\textcolor{red}{$\full$}), the two outer layers in phase and the middle layer in anti-phase (\textcolor{red}{ $\chain$}), and finally no density modulation in the middle layer and the outer layers in anti-phase (\textcolor{red}{$\dashed$})
(b) Classical interaction energy between two layers for the in-phase ($\full$) and anti-phase (\textcolor{green}{$\dashed$}) configurations as a function of the wavelength of the density modulations.}\label{fig:Splitting}
\end{center}
\end{figure}

The same result holds for more than two layers: The mode with the stripes in phase always becomes unstable first, irrespective of the 
value of $\theta$, except for $\theta=\pi/2$ where the modes are degenerate. 
 This is illustrated in figure~\ref{fig:Splitting}(a) which depicts the critical coupling strength for the even and odd modes for the cases of two and three layers. The modes with higher $g_c$ were calculated using the self-energy at the value for the lowest modes for simplicity. This changes the obtained values slightly but retains the relative ordering of the modes.
For the case of three layers, there is an additional mode with no density fluctuations in the middle layer. 
It goes unstable for a coupling strength in between the values for the even 
and odd modes. This agrees with the results found within the RPA~\cite{Zinner}~\footnote{The odd mode and the mode with 
no density modulations in the middle layer was mistakenly swapped around in the text of ref.~\cite{Zinner}.}.

\section{Classical model}
We now demonstrate that the effects of neighboring layers can be understood from simple classical considerations. First, it is easy to show, see~(\ref{eq:classicalplane}), that the classical interaction energy of a single dipole with an infinite layer 
of dipoles with homogenous density is zero. Second, we analyze the interaction between stripes by calculating the classical interaction energy between a single dipole and stripes of increased/decreased density of dipoles in a layer separated by a distance $d$. As explained in detail in \ref{app:Classical}, a straightforward calculation gives that this interaction 
energy is 
\begin{equation}
 \label{eq:Vclassical}
 V_{\rm classical}=\mp\frac{2\pi^2D^2\gamma\cos^2\theta}{{\lambda_c^2}}\big[\csch^2(\pi d/\lambda_c)+\sech^2(\pi d/\lambda_c)\big],
\end{equation}
where the $-$ is for the in-phase case where one stripe in the plane is directly above the single dipole, and $+$ is for the anti-phase case where 
the projection of the dipole onto the plane is in between two stripes. The linear density of dipoles within a stripe is given by $\gamma$, and 
$\lambda_c$ is the distance between the stripes within the layer. Equation (\ref{eq:Vclassical}) is plotted in figure~\ref{fig:Splitting}(b).

This demonstrates that the in-phase/out-of-phase configuration of the stripes has a negative/positive interaction energy 
 thereby explaining why they decrease/increase the critical coupling strength $g_c$ for stripe formation as compared 
to the single layer case. The $\cos^2\theta$ dependence in (\ref{eq:Vclassical}) furthermore explains why the effect decreases for increasing angle, with the modes 
becoming degenerate for $\theta=\pi/2$ as seen in figure~\ref{fig:Splitting}(a). It is reassuring that our full quantum mechanical calculation which is rather numerically involved, agrees with this classical analysis.
 
\section{Experimental realisations}
We now examine the importance of the interlayer correlations discussed above for typical experiments. 
 In an experiment where the planar confinement is caused by a 1D optical lattice formed by counter propagation lasers with wavelength $\lambda$, the layer distance will be $d=\lambda/2$. Figure~\ref{fig:ExpFig} shows the critical dipole strength times the square root of the mass, $\sqrt mp$, as a function of density for one, two, and three layers and the dipoles perpendicular to the planes for a typical wavelength of $\lambda=1064\, {\rm nm}$. For comparison, in \cite{miranda2011} the JILA group reports a peak density $\sqrt{n_{\rm 2D}}=0.58\, \mu{\rm m}^{-1}$ or $d\kfo=1.1$ in an experiment with $^{40}$K$^{87}$Rb molecules in the rotational and vibrational ground state. In the figure, the horisontal lines are the permanent electrical dipole moment times the square root of the mass for several experimentally relevant fermionic dipolar molecules. Note that the effective dipole moment in the trap is somewhat smaller, since the dipoles are not aligned perfectly. It is however of the same order of magnitude as the permanent dipole moment; in~\cite{ni2010} the JILA group reports a maximum value for the average dipole moment in the laboratory frame of about 40$\%$ of the permanent moment.

From the figure, we see that the molecules $^7$Li$^{40}$K, $^{23}$Na$^{40}$K, and $^{40}$K$^{137}$Cs and moreover $^{6}$Li$^{87}$Rb and $^{6}$Li$^{133}$Cs, 
which lie outside the figure, have such large values of $\sqrt mp$, that one will observe stripe formation already in the regime of relatively low density where multilayer effects are important. This demonstrates the experimental relevance of the results discussed in this paper. 
The $^{40}$K$^{87}$Rb and $^6$Li$^{23}$Na molecules on the other hand have small values of $\sqrt m p$, and the density required to observe stripe formation is so high that interlayer correlations are not important. %These are all the fermionic molecules consisting of two different alkali metals.

The molecules containing Li are all chemically unstable \cite{zuchowski2010} in the sense that the reaction $2{\rm YX}\to {\rm Y}_2+{\rm X}_2$ is exothermic for 
Y=Li and X any other alkali metal. Molecules of $^{23}$Na$^{40}$K and $^{40}$K$^{137}$Cs are however chemically stable in this sense, so they are prime candidates for studying the density wave instability.

\subsection{Experimental issues and detection}
Experiments with polar molecules operate at finite temperatures. The JILA experiments with KRb have reported temperatures 
down to $T=220$ nK or $T/T_F=1.4$ \cite{wang2010}. This is close to but not quite in the 
degenerate regime, so a decrease in temperature is most likely needed to see many of the 
predicted phases. Furthermore, in the low-dimensional setups created by optical lattices, heating can occur as 
the lattice is turned on, demanding that the initial temperature be even lower to reach critical temperatures.

In the strict 2D limit, the molecules are completely confined in the transverse direction. This means that their motion is reduced to a strict planar geometry. In addition, we do not allow any tunneling between the layers. The layer index can be considered an effective spin label. In this case, the
non-zero temperature phase transition to ordered states such as the density wave is governed by the Berezinskii-Kosterlitz-Thouless (BKT) transition \cite{bkt1,bkt2}, and no true long-range order occurs. One consequence is that the BKT transition temperature is below the transition temperature obtained from mean-field theory.

However, studies of one-dimensional arrays of tubes with two-component fermions that can undergo a pairing transition indicate that weak tunneling between the tubes can stabilize long-range order \cite{parish2007}. We imagine that this could also work for our multilayer system, i.e. by allowing weak tunneling between the layers the density-wave ordering becomes long-range and the transition temperature may be increased from the BKT value to the mean-field prediction. This intriguing possibility will be explored in future work.

For detection of the density wave state in the multilayered setup a number of different techniques are possible. A quantum non-demolition measurement \cite{wunsch2011,mekhov2007,eckert2008,vega2008,mekhov2009,bruun2009,zinner2011b} 
could be used to detect the large density fluctuations close to the phase transition. Alternatively, light scattering experiments proposed to detect dimerized pairing in multilayers \cite{potter2010} could be adapted to the density wave case. 

\section{Conclusions} 
Using a conserving Hartree-Fock approximation, we examined the density wave instability of aligned fermionic dipoles 
moving in equidistant planes. We found that while exchange correlations suppress the instability, it can be significantly enhanced 
by correlations between the layers. The inter-layer correlations exhibit 
an interesting dependence on the dipolar angle, and they result in the density waves
in the different layers to be in-phase for all angles. We furthermore demonstrated that the physics of the interlayer correlations 
can be understood from a classical model. The density wave instability was shown to be experimentally accesible with realistic densities 
for experiments using $^7$Li$^{40}$K, $^{6}$Li$^{87}$Rb, $^{6}$Li$^{133}$Cs, $^{23}$Na$^{40}$K and $^{40}$K$^{137}$Cs molecules. 
For these molecules, interlayer correlations were furthermore predicted to decrease the critical coupling strength for density waves significantly.

\section*{Acknowledgments}
We thank M. Baranov, L. Sieberer and M. Babadi for discussions about their work, and D.-W. Wang for discussion about the dipolar interaction in layered geometries. This work is supported by the Danish Council for Independent Research | Natural Sciences.

\appendix

\section{Numerics\label{app:Numerics}}
\subsection{Green's function \label{app:GreenNum}}
Since we use the conserving approximation, the limits on the self-energy integral in \eqref{eq:SE} is dependent on $\Sigma$ itself. Since we only consider $T=0$, the ground state of the system is described by a deformed Fermi sea. Inspired by \cite{Sieberer}, we make a Fourier series for the function $h(\phi)=k_F^2(\phi)/{\kfo}^2-1=\sum_{n=1}^6 a_n\cos(2n\phi)$ keeping the first six terms, which gives the deformation relative to the non-interacting Fermi sea, and determine the coefficients for a given interaction strength by an iterative procedure. An alternative is to approximate the Fermi sea by an ellipse and do a variational calculation as derived in \cite{HanPhysRevA.77.061603} and applied in \cite{Bruun,Yamaguchi}. Both approaches have been implemented by the authors of this paper, and the results are very similar. Interested readers can find the details of the procedure in \cite{Sieberer}.

\subsection{Particle-hole scattering matrix}
The particle-hole scattering matrix is determined by \eqref{eq:scat}, where the second variable $(\kpv)$ is not integrated, so it just appears as a parameter. 

The Fermi functions in \eqref{eq:scat} restrict the two dimensional integration domain to the (deformed) Fermi sea and a Fermi sea displaced by $\qv$. In order to reduce the numerical complexity, we define the functions $\Gamma_{ij}^{\pm}(\kv,\kpv,q)=\Gamma_{ij}(\kv,\kpv,q)\pm\Gamma_{ij}(-\kv-\qv,\kpv,q)$. The inversion symmetry of both inter- and intralayer potentials $V(\kv)=V(-\kv)$ gives $f(-\kv)=f(\kv)$ and $\epsilon(-\kv-\qv)=\epsilon(\kv+\qv)$ which allows for a shift of the displaced Fermi sea to give for the static ($\omega=0$) scattering matrix
\begin{align}
\fl \Gamma_{ij}^{\pm}(\kv,\kpv,\qv)=&I_{ij}^{\pm}(\kv,\kpv,\qv)+\sum_{\kppv l}I_{il}^{\pm}(\kv,\kppv,\qv)\frac{f_{l\kv''}}{\epsilon_{l\kv''}-\epsilon_{l\kv''+\qv}} \Gamma_{lj}^{\pm}(\kppv,\kpv,\qv),\label{eq:scatpm}
\end{align}
where $I_{ij}^{\pm}(\kv,\kpv,\qv)=I_{ij}(\kv,\kpv,\qv)\pm I_{ij}(-\kv-\qv,\kpv,\qv)$. The scattering matrix is then given by $\Gamma(\kv,\kpv,\qv)=\frac{1}{2}[\Gamma^{+}(\kv,\kpv,\qv)+\Gamma^{-}(\kv,\kpv,\qv)]$ and thus has poles if and only if at least one of the symmetrized $\Gamma$'s has a pole (unless the poles of $\Gamma^{+}$ and $\Gamma^{-}$ happen to cancel).
 
The Fermi sea is deformed by the dipole-dipole interaction, so we write the sum as an integral in polar coordinates and for fixed $\phi''$ parametrize the norm integral by $x''=k''/k_F(\phi'')$. Let $\hat{\kv}$ be unit vector in the direction of $\kv$, then the integral becomes
\begin{align*}
\fl& \sum_{\kppv l}I_{il}^{\pm}(\kv,\kppv,\qv)\frac{f_{l\kv''}}{\epsilon_{l\kv''}-\epsilon_{l\kv''+\qv}} \\
\fl &=\frac{1}{2\pi}\sum_{l}\int_{0}^{2\pi}\frac{\rmd\phi''}{2\pi}\int_0^{1}\rmd x'' I_{il}^{\pm}\big(\kfo\sqrt{1+h(\phi)}x \hat{\kv},\kfo\sqrt{1+h(\phi'')}x'' \hat{\kv}'',2\kfo\sqrt{1+h(\pi/2)} \hat{\qv}\big)\\
\fl&\cdot \frac{\kfo\sqrt{1+h(\phi'')}x''}{\epsilon_l\big(\kfo\sqrt{1+h(\phi'')}x''\hat{\kv}''\big)-\epsilon_l \big (\kfo\sqrt{1+h(\phi'')}x''\hat{\kv}''+2\kfo\sqrt{1+h({\pi}/{2})}\hat{\qv}\big)}
\end{align*}

An approach to solving the integral equation \eqref{eq:scatpm} for $\Gamma_{ij}^{\pm}$ is to choose suitable abscissa $\kv_{\alpha}$ and weights $w_{\alpha}$ for points in the single Fermi sea and approximate the integral by a sum. Then \eqref{eq:scat} becomes a matrix equation where $(i,j)$'th block of the matrices describes the interaction on layer $i$ from layer $j$.
 We suppress the common $q$-dependency and introduce the weights as a diagonal matrix $W$. In the double indices of layer number (roman letter) and $\kv$-grid point (Greek letter) the equation reads
\begin{align}
&\Gamma_{i\alpha,j\beta}^{\pm}=I_{i\alpha,j\beta}^{\pm}+I_{i\alpha,l\gamma}^{\pm}W_{l\gamma}\cf_{l\gamma}\Gamma_{l\gamma,j\beta}^{\pm} \quad \mathrm{or}\nonumber\\
&\big[\delta_{i\alpha, l\gamma}-I_{i\alpha,l\gamma}^{\pm}W_{l\gamma}\cf_{l\gamma}\big] \Gamma_{l\gamma,j\beta}^{\pm}=I_{i\alpha,j\beta}^{\pm} \label{eq:scatmat},
\end{align}
where the diagonal matrix $\cf$ has entries $\cf_{l,\gamma}=\frac{\abs{\kv_{\gamma}}}{\epsilon_{l\kv_{\gamma}}-\epsilon_{l\kv_{\gamma}+\qv}}$. 

The irreducible interaction is finite, so the scattering matrix diverges when the matrix in the brackets becomes singular. This happens when the matrix $I_{i\alpha,l\gamma}W_{l\gamma}\cf_{l\gamma}$ has an eigenvalue of $1$ (see also \ref{app:nearcrit}). For $I^{-}$ the direct interaction cancels and the layers decouple as there is no exchange between different layers. For the single layer we find that the matrix has only negative eigenvalues, so $\Gamma^-$ never contributes to the divergence.

For fixed $\phi_c=\pi/2$ and $q=2k_F(\phi_c)$ we vary $g$ until the first eigenvalue crosses 1. The difference of the single particle energies of the particle and hole in the denominator of $\cf$ gives rise to an integrable singularity at the edge of the integration region at $\phi''=3\pi/2,k''=2k_F(\phi_c)$, so we partition the $\phi''$ interval and cast 60 points in the $[-\pi/2+\Delta\phi,3\pi/2-\Delta\phi]$ and 10 points in $[3\pi/2-\Delta\phi,3\pi/2+\Delta\phi]$. For the $x=k''/k_F(\phi)$ variable we cast 10 points in $[0,1-\Delta x]$ and 30 points in $[1-\Delta x,1]$.
We choose the abscissa and weights in the four intervals by a Gauss-Legendre quadrature rule. For $\Delta \phi =0.05$ and $\Delta x=0.02$, a tripling of the number of points in either interval gives a change of the critical coupling strength that is less than the absolute tolerance $2\cdot 10^{-3}$.

The matrix depends on $g$ explicitly from the potential in $I$, but also indirectly through the self energies and Fermi function. The iterative procedure is as follows: For the current guess for $g_c$, calculate the deformation $f(\phi)$ of the Fermi sea according to \ref{app:GreenNum}, rescale the $k$-points and calculate the matrix $\frac{I_{i\alpha,l\gamma}^{\pm}}{g}W_{l\gamma}\cf_{l\gamma}$, find the largest eigenvalue $\lambda_1$ and set the new guess to $g=\lambda^{-1}$. The iteration is terminated when the absolute change in $g$ is less than $2\cdot 10^{-3}$.

\section{Scattering matrix near the critical point \label{app:nearcrit}}
As $g\to g_c$ for fixed $\qv=\qv_c$ the matrix in the square brackets in \eqref{eq:scatmat} becomes singular and the response blows up. If we make a spectral decomposition of $I^{\pm}_{i\alpha,l\gamma}W_{l\gamma}\chi^f_{l\gamma}$, the eigenvector with eigenvalue 1 will dominate in the vicinity of the divergence. To see this (for simplicity write $I=I^{\pm}_{i\alpha,l\gamma}$ and for the diagonal matrix $W_{l\gamma}\chi^f_{l\gamma}=D$), consider a $g<g_c$, so that $ID$ has eigenvalues $\lambda_i<1 \forall i$. We need to find the inverse $(\Ide -ID)^{-1}$, so we make an eigenvalue decomposition for $ID$, say $ID=VAV^{-1}$ with $A={\rm diag}(\lambda_1,\dots)$. Then $(ID)^n=VA^nV^{-1}\to 0$ for $n\to \infty$. Now take the geometric series $\sum_{n=0}^{\infty}(ID)^n=V\sum_{n=0}^{\infty} A^n V^{-1}=V{\rm diag}[1/(1-\lambda_1),\dots]V^{-1}$. So we have an inverse, as
\begin{equation}
 (\Ide-ID)\sum_{n=0}^{\infty}(ID)^n=\Ide-\lim_{n\to\infty}(ID)^n=\Ide.
\end{equation}
Inserting in \eqref{eq:matstruc}
\begin{equation}\label{eq:gaminv}
\Gamma=V{\rm diag}[1/(1-\lambda_1),\dots]V^{-1}I
\end{equation}
As $g\to g_c$, the first $1/(1-\lambda_1)$ is divergent.

\subsection{Bilayer }
As mentioned above, the $\Gamma^{-}$ part is not divergent, so in the vicinity of the critical point, the scattering matrix is determined by the behaviour of $\Gamma^{+}$, $\Gamma\approx\frac{1}{2}\Gamma^{+}$.
For the bilayer case we write the matrix structure of the $2n\times 2n$ block matrix equation \eqref{eq:scatmat} explicitly as: the identity matrix $\delta_{\alpha,\gamma}=\Ide$, the symmetric block matrices of the irreducible particle-hole interaction $I^{+}_{1\alpha,1\gamma}(\qv=\qv_c)=I^{+}_{2\alpha,2\gamma}(\qv=\qv_c)=I^{+}_{1}$, $I^{+}_{1\alpha,2\gamma}(\qv=\qv_c)=I^{+}_{2\alpha,1\gamma}(\qv=\qv_c)=I^{+}_{2}$, the diagonal product of the weights and particle-hole propagator $w_{\alpha}\cf_{\alpha}(\qv=\qv_c)\delta_{\alpha,\gamma}=D$ and the particle-hole scattering matrix $\Gamma^{+}_{l\gamma,j\beta}(\qv=\qv_c)=\Gamma^{+}_{l,j}$.
\begin{widetext}
\begin{align}
\fl
\bigg[
 \begin{qmatrix}
\Ide&0\\
0&\Ide
 \end{qmatrix}
-
\begin{qmatrix}
I^{+}_{1} & I^{+}_{2}\\
I^{+}_{2} & I^{+}_{1}\\
\end{qmatrix}
\begin{qmatrix}
D&0\\
0&D
\end{qmatrix}
\bigg]
\begin{qmatrix}
\Gamma^{+}_{1,1} & \Gamma^{+}_{1,2}\\
\Gamma^{+}_{2,1} &\Gamma^{+}_{2,2}
 \end{qmatrix}
=
 \begin{qmatrix}
I^{+}_{1} & I^{+}_{2}\\
I^{+}_{2} &I^{+}_{1}
 \end{qmatrix}\label{eq:matstruc}
\end{align}
\end{widetext}
For $g<g_c$ we have the inversion in (\ref{eq:gaminv}) above and now wish to examine the diagonalizing matrix $V$. The matrix $I^{+}D$ has the symmetry found by switching blocks by multiplying with $C=\begin{bmatrix} 0 & \Ide\\ \Ide & 0 \end{bmatrix}$: $C$ has $C^{-1}= C$ and commutes with both $I^{+}$ and $D$ since all 3 are (block) symmetric, thus it commutes with the product. $C$ is it's own inverse, so it has eigenvalues $\pm 1$. Each has $n$ linearly independent eigenvectors of the form $v=[w^T,\pm w^T]$ for $\lambda=\pm 1$ respectively. Because the matrices commute, the eigenvectors of $I^{+}D$ can be chosen to have the same form and the numerics show that the eigenvector with the largest eigenvalue has the $+$-form. The diagonalizing matrix can be thus chosen to be of the block form
\begin{equation}
 \label{eq:V}
V=\begin{bmatrix} W_1 & W_2\\ W_1 & -W_2 \end{bmatrix} \quad {\rm with} \quad V^{-1}=\frac{1}{2}\begin{bmatrix} W_1^{-1} & W_1^{-1}\\ W_2^{-1} & -W_2^{-1} \end{bmatrix},
 \end{equation}
where the $n\times n$ matrices $W_i$ are determined by the eigenvectors.
As $g\to g_c$ only the first $1/(1-\lambda_1)$ is divergent in the decomposition \eqref{eq:gaminv}. To find the response functions, we need to sum all matrix elements in each block \eqref{eq:densityresponse}. Near the divergence all but the first entry in the diagonal matrix can be ignored, so
\begin{equation}
\Gamma\approx\frac{1}{2}\Gamma^{+}\approx
 \begin{b3matrix}
\frac{1}{2(1-\lambda_1)}\vec{v}_1&0&\dots
 \end{b3matrix}
V^{-1}I,
\end{equation}
where $v_1=\frac{1}{2}\begin{b1matrix}w_1 \\ w_1\end{b1matrix}$, and $w_1$ is the first column of $W_1$ from (\ref{eq:V}). From \eqref{eq:V} we see, that the first row of $V^{-1}$ has the structure $y_1^{T}=\frac{1}{2}[z_1^T,z_1^T]$, where $z_1^T$ is the first row of $W_1^{-1}$. The matrix structure is now
\begin{widetext}
\begin{align}
\begin{bmatrix}
\Gamma_{1,1} & \Gamma_{1,2}\\
\Gamma_{2,1} &\Gamma_{2,2}
\end{bmatrix}
&\approx\frac{1}{4(1-\lambda_1)}
\begin{bmatrix}w_1z_1^T(I^{+}_1+I^{+}_2) & w_1z_1^T(I^{+}_1+I^{+}_2) \\w_1z_1^T(I^{+}_1+I^{+}_2) & w_1z_1^T(I^{+}_1+I^{+}_2) \end{bmatrix}\label{eq:scatmatdiv}
\end{align}
\end{widetext}
We see that the blocks of the scattering matrix are all the same close to the divergence $\Gamma_{ij}=\Gamma_c$. The response function is given by \eqref{eq:densityresponse}, so $\chi_{i,j}$ is found by multiplying the block matrix in \eqref{eq:scatmatdiv} by the diagonal matrix $\cf$ on both sides and then summing all matrix elements within each $n\times n$ block and finally adding $\chi^0$, the response function for non-interacting dipoles with HF single particle energy.
The $g$-dependency of the $I^{+}D$-matrix comes both directly from the dipole-dipole interaction in $I^{+}=g\tilde{I}^{+}$ where $\tilde{I}^{+}$ is determined by the vectors $\kv,\kppv,\qv$ and from the self-energy in $\cf$.

The critical $g_c$ is defined by the nonlinear eigenvalue equation $g(\tilde{I^{+}})D(g)v=\lambda_1 v$, so by expanding near $g=g_c$ we see $1-\lambda \propto g_c-g$, so after the summation, the $2\times 2$ structure is 
\begin{align}
\label{eq:densityresponse2}
\begin{bmatrix}
\chi_{1,1} & \chi_{1,2}\\
 \chi_{2,1} & \chi_{2,2}
 \end{bmatrix}
=&
\begin{bmatrix}
\chi^0 & 0\\
0& \chi^0
\end{bmatrix}
+\frac{1}{1-g/g_c}
\begin{bmatrix}
\chi^c & \chi^c \\ 
\chi^c& \chi^c
 \end{bmatrix}.
\end{align}

The numerics shows that the second largest eigenvalue value has an eigenvector of the type $v_2^T=[w_2^T, -w_2^T]$. Near the corresponding $g$, the inversion formula for $\{1-(I^{+}D)\}$ does not work since $(I^{+}D)^n$ does not go to zero. This is because the system has already gone unstable. If we ignore this fact, say if the highest eigenvalue actually was the one with this eigenvector, the rank one approximation to the inverse would be the outer product of the $(n+1)$th column and row vectors of $V$ and $V^{-1}$ from \eqref{eq:V}, $y_2^T=\frac{1}{2}[z_2^T,-z_2^T]$ so the response near this $g_{c2}$ would be
\begin{align}\label{eq:dens2nd}
\begin{bmatrix} \chi_{1,1} & \chi_{1,2}\\ \chi_{2,1} &\chi_{2,2} \end{bmatrix}
=\begin{bmatrix}
\chi^0 & 0\\
0& \chi^0
\end{bmatrix}
+\frac{1}{1-g/g_c}
\begin{bmatrix}
\chi_2^c & -\chi_2^c \\ 
-\chi_2^c& \chi_2^c
 \end{bmatrix}.
\end{align}.

\section{Classical calculations \label{app:Classical}}
First we show that a dipole in one layer does not feel a homogeneous distribution of dipoles in another (infinite) layer. The dipoles are in the $xz$-plane and aligned by the external field as $\pw=D(\sin\theta_E,0,\cos\theta_E)$ and the two layers are separated by a distance $d$. The angle between the relative vector and the dipole orientation is 
\begin{equation}
\cos^2\theta_r=\frac{(\hat{p}\cdot \rw)^2}{r^2}=\frac{(x\sin\theta_E+d\cos\theta_E)^2}{r^2},
\end{equation}
while dipole-dipole interaction is given by \eqref{eq:Vrealspace}. The calculations is done in a polar coordinate system in the $xy$-plane, $\rho^2=x^2+y^2$, so
\begin{align}\label{eq:classicalplane}
\fl V_{\rm plane}&=D^2\ntd \int_0^{\infty}\rmd \rho \int_0^{2\pi} \rmd\phi \rho \bigg (\frac{1}{\big(\rho^2+d^2\big)^{3/2}}-3\frac{(\rho\cos\phi\sin\theta_E +d\cos\theta_E)^2}{\big(\rho^2+d^2\big)^{5/2}} \bigg)\nonumber\\
\fl&=0
\end{align}
This shows that any homogeneous background density cancels, so the interaction with a secondary layer is only dependent on the deviation from the average density. A density lower than the background can be modelled by changing the sign of the potential.

Since the homogeneous background density does not contribute to the interlayer interaction, we can model the density wave phase by a periodic series of lines with alternating sign on the dipole-dipole interaction. The distance (wavelength of the density wave) between two lines with the same change in density is denoted $\lambda_c$. The full quantum mechanical calculation~(\ref{eq:biresponsepole}) shows that the collective eigenmodes correspond to in-phase and anti-phase density modulations in the two layers, so we only consider these two extremes.

We calculate the interaction between a single dipole in a stripe in one layer and the stripes in the other layer by splitting the lines into two sub-series. The first has lines directly on top while the lines of the other series are shifted by $\lambda_c/2$. For the in-phase density modulations the first sub-series has excess density while the second has decreased density. For the anti-phase configuration the signs of the interactions are switched, but the geometry is otherwise the same. Using the coordinate system from figure~\ref{fig:setup}, the stripes are parallel to the $x$-axis. The relative coordinates between the dipole and a point in line $n$ in the first sub-series and second sub-series is
\begin{equation}
\rw_{1,n}=[x,\lambda_cn,d] \quad \rw_{2,n}=[x,\lambda_c(n-1/2),d],
\end{equation}
respectively.
For convenience we define $a_n^2=n^2\lambda_c^2+d^2$ or $a_n^2=(n+\frac{1}{2})^2\lambda_c^2+d^2$ as the squared distance in the $yz$-plane in the two cases. The dipoles are assumed to be smeared out along the stripes with a linear dipole density of $\gamma$. So the interaction with line $n$ is
\begin{align}
V_{n}&=D^2\gamma\int_{-\infty}^{\infty}\rmd x \frac{1-3\cos(\theta_r)^2}{r^3}
=2D^2\gamma\cos^2\theta_E\frac{a_n^2-2d^2}{a_n^4}
\end{align}
Summing the contributions from all stripes in both sub-series gives the interaction between a single dipole in a stripe with all the stripes in the other layer
\begin{equation}
 \label{eq:Vclassicalap}
 V_{\rm classical}=\mp\frac{2\pi^2D^2\gamma\cos^2\theta_E}{{\lambda_c^2}}\big[\csch^2(\pi d/\lambda_c)+\sech^2(\pi d/\lambda_c)\big],
\end{equation}
where the $-$ is for the in-phase and $+$ is for the anti-phase configuration.

\end{document}